\documentclass[cameraready]{Interspeech}
\usepackage[table,xcdraw]{xcolor}
\usepackage{multirow}
\usepackage{caption}
\usepackage{subcaption}

\title{Sexualised synthetic personas encode and amplify gendered power asymmetries through voice}

\author[affiliation={1}, orcid=0009-0006-9871-4475, equalcontribution, correspondingauthor]{Alice}{Ross}
\author[affiliation={1}, orcid=0009-0000-4721-9147, equalcontribution, correspondingauthor]{Ariadna}{Sanchez}
\author[affiliation={2}]{Elin}{Kanhov}
\author[affiliation={1}]{Catherine}{Lai}
\author[affiliation={2}]{Éva}{Székely}

\address{
    $^1$ University of Edinburgh, United Kingdom \\
    $^2$ KTH Royal Institute of Technology, Sweden
}

\email{alice.ross@ed.ac.uk, ariadna.sanchez@ed.ac.uk}


\usepackage{comment}

\begin{document}

\maketitle

\begin{abstract}
This work examines sexualised AI-generated English-speaking voices offered by a popular commercial platform. 
New technologies may enable sexual empowerment and greater diversity in gender expression, yet toxic masculinity, heteronormativity, and the abuse of women and LGBTQ+ people remain pervasive online.
Drawing on a Feminist HCI perspective, we examine how commercial voice AI systems reproduce and circulate particular performances of gender. We conducted a listening experiment with a diverse group of  listeners, combining quantitative adjective selection, qualitative free-text responses, and acoustic analysis. Participants evaluated male- and female-coded voices presented with either sexualised scripts or neutral text. Results reveal a narrow range of gender expression, largely binary and heteronormative. Female-coded voices are more frequently described using sexualised and submissive terms, while male-coded voices are more often associated with dominance and positive traits.

\end{abstract}

\section{Introduction} \label{intro}

The internet and novel technologies can provide supportive spaces for marginalised people to create connections and perform identity. While they can offer opportunities for women's sexual empowerment \cite{mondin2017tumblr, mckeown2018my}, or for queer and trans people to experiment with sexuality and gender (e.g., \cite{oriordan2007queer, haimson2021tumblr}), harassment and victimisation of these groups persist as common features of internet discourse \cite{curry2024subjective, twomey2025what}.
Concurrently, text-to-speech (TTS) (also recently called `Voice AI') has been subjected to gendering and social perceptions. Humans make social judgments about the appearance, characteristics and personality traits of the imagined `speakers' behind TTS voices \cite{de2025sketching, lilley2024social, ross2026sound}. For this reason, in the 2010s, the characterisation of early voice assistants (VAs) like Alexa and Siri as `female by default', conforming to a stereotypical `secretary' persona, rightly drew criticism in academic and popular press \cite{west2019id, abercrombie2021alexa, weisman2020instantaneity}. Now, we turn our attention to a newer arena and find evidence that some of the same outdated roles may persist. The important difference is in the nature of the product: the promise of voice assistants was to reduce friction in user interactions with existing functions (e.g., setting alarms) while in commercial voice AI platforms, the human-like generated voice itself is offered for sale. Users are encouraged to play Pygmalion, prompting and personalising the TTS output to their precise specifications and then telling it what to say, with the option of adding embodied paralinguistic features like giggles and sighs.

In 2025 (still visible via the internet archive),
the TTS section of ElevenLabs' website presented various categories of `emotional' TTS voices, with the top three including \textit{seductive} and \textit{sultry}. These categories contained lists of voices, almost always clearly coded as female and with feminine names (e.g., Natasha).
These categories still exist, but their links are now less prominent on the home page and several are now populated by prompt-generated voices with generic names like ‘The midnight enchantress’. The categories \textit{flirty}, \textit{attractive}, \textit{submissive}, etc, contain a list of these personas with editable prompts and sample audio, evenly split (2+2) between male and female coded (we encountered one gender neutral voice in our initial exploration, representing an androgynous, ageless being: `The abyssal echo'). Ostensibly, then, ElevenLabs now offers men's and women's voices on an equal footing, with similar prompts and preset flirtatious messages. Regardless of this `gendered parity', it is unknown whether these voices are perceived similarly by users, whether they embody specific portrayals of gender and power, and whether they reinforce harmful stereotypes.

In this work, we aim to explore listeners' reactions to particular speech styles (defined by ElevenLabs as \textit{flirt, flirty,} and \textit{temptress}) within sets of synthesised, \textit{gendered} voices. We consider them gendered in that a sex/gender term (in this case, \textit{male} or \textit{female}) is included in the prompt. 
We approach these voices, as representative outputs of a larger system, from the perspective of Feminist HCI \cite{rode2011theoretical, dignazio2023data}, drawing attention to how technologies are shaped by and encode gender relations. 

To explore perceptions of the voices, we carry out a novel listening test, gathering reactions from a diverse set of North American English-speaking listeners. We first examine whether male- and female-coded voices differ in the overall distribution of evaluative judgments assigned by listeners. We then distinguish between effects of vocal and prosodic characteristics and those of linguistic content, by comparing responses across conditions where voice style and spoken text are independently varied.
Finally, we explore whether listener characteristics, including gender identity and reported attraction, are associated with differences in perception of the voices.
Analysing the resulting quantitative, qualitative and acoustic data, we find that a narrow range of gender expressions, almost exclusively binary and heteronormative, are represented. 
Listener evaluations reveal systematic asymmetries between male- and female-coded voices: male voices are more often described using positive and dominant adjectives, whereas female voices are more frequently characterised as submissive and sexualised. These differences persist even when the spoken text is held constant, suggesting that stylised prosodic and paralinguistic features in the female-coded voices contribute to these perceptions.

\section{Related work}

Machine learning models, especially natural language models using word embeddings, reproduce and amplify stereotypes present in their training data (e.g., \cite{garg2018word}).
In the study of human speech, listeners have been found to make different judgments about information conveyed by voice quality in (perceived) male vs female speakers (e.g., \cite{pearsell2023effects}; female speakers were rated more negatively for some personality traits when producing \textit{creaky} voice, while \textit{smiling} voice recordings were consistently rated higher for women compared to men).
In the study of anthropomorphic TTS systems, Holliday explored listeners' judgments of social and personal traits in US English Siri voices \cite{holliday2023siri}. Users' subjective perceptions of specific qualities indexed in TTS voices, such as ethnicity \cite{pinhanez2024creating}, non-binary gender expression \cite{hope2025voices}, and kawaii \cite{mandai2025super}, have been investigated. Along this line, we consider perceptions of attractiveness and personal agency or dominance. There exists extensive research on perceptions of these qualities from human speech (e.g., \cite{hodges2010different, borkowska2011female, re2012preferences, quene2016attractiveness}). However, we are not aware of any existing work exploring this in AI-generated voices, particularly studies that consider listeners with diverse sexual orientations.

\section{Listening experiment}

\subsection{Stimuli} 
\label{sec:stimuli}

Participants are presented with audio samples of different voices generated with ElevenLabs' `Voice Library'\footnote{\url{https://ariadnasc.github.io/synth-personas}}. Some voices are advertised with sexualised adjectives, e.g., \textit{flirty}, \textit{flirt} or \textit{temptress}, while others are advertised with non-sexualised adjectives, e.g., \textit{informative}, \textit{presenter}, or \textit{educational}. For these adjectives, ElevenLabs only advertises `female' and `male' voices. We select the most popular `female' and `male' voice in each category, resulting in the following list of voice personas: \textit{The Parisian temptress} (flirt; female), \textit{The Southern gentleman} (flirt; male), \textit{The sophisticated charmer} (flirty; female), \textit{The velvet rogue} (flirty; male), \textit{The midnight enchantress} (temptress; female), \textit{The velvet gentleman} (temptress; male), \textit{The tech expert} (informative; female), \textit{The documentary narrator} (informative; male), \textit{The professional news anchor} (presenter; female), \textit{The science enthusiast} (educational; male).

For the voices with sexualised adjectives, we synthesise sexualised and informative texts to isolate the effect of voice and linguistic content. As sexualised text, we use those presented to users on ElevenLabs' website. As informative text, we synthesise various excerpts of a modified version of the Rainbow passage \cite{dietsch2023revisiting}. As a baseline, we synthesise the same excerpts of the Rainbow Passage for the voices with non-sexualised adjectives (i.e., informative, presenter, educational). In total, we generate one sample of sexualised text per sexualised voice (6 in total), and two samples of Rainbow text for both sexualised and non-sexualised voices (12 in total for each type of voice). This design enables controlled comparisons of voice style independent of linguistic content while keeping the length of the study to 20 minutes.

\begin{figure*}[t]
    \centering
    \includegraphics[width=\linewidth, trim = 50 230 50 200]{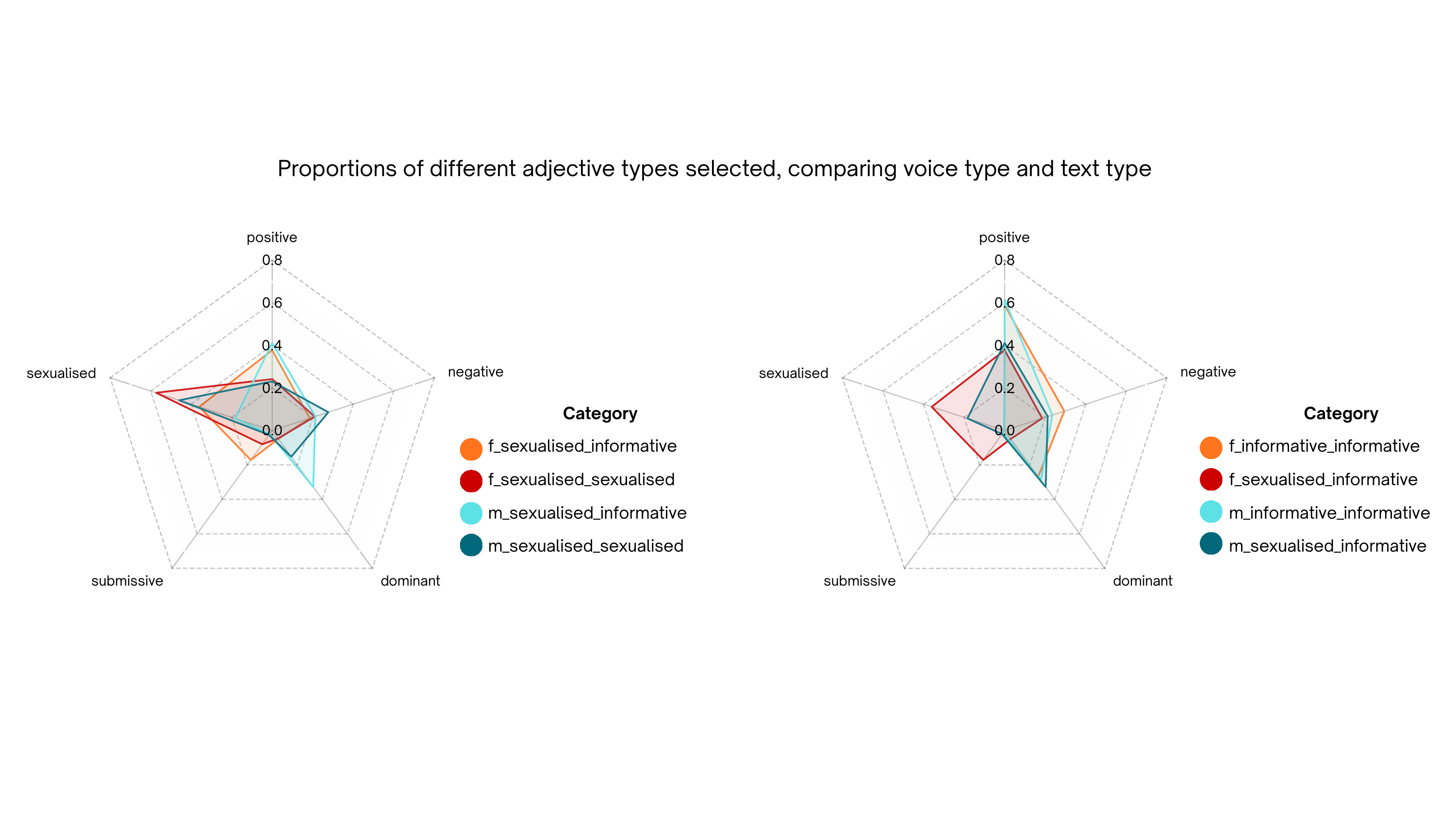}
    \caption{Comparing the proportions of positive, negative, dominant, submissive, and sexualised adjectives given to different sets of male and female voices. Blue/turquoise = `male' voices; red/orange = `female' voices.}
    \label{fig:spider_plots_combined}
    \vspace{-1em}
\end{figure*}

\subsection{Method}
In the listening test, participants listen to the audio samples and select the three most suitable adjectives from a list to describe the voice. The assessment of language attitudes by selecting adjectives from balanced lists is established in sociolinguistics \cite{hall2015tourists}, which we apply to the study of voice characteristics. We adopted this task rather than Likert-type rating scales commonly used in TTS evaluation, or exclusively free-text responses, as it allows us to collect a large number of evaluative judgments while minimising participant fatigue. In addition, this approach avoids variability in scale interpretation across participants. Compared to fully open-ended responses, a controlled adjective list also facilitates systematic comparison across participants while still capturing subtle variation in perception per stimulus. The listening test was presented to participants as a website created using the jsPsych toolkit \cite{de2015jspsych}.

Participants complete a total of 30 trials with the stimuli explained in section \ref{sec:stimuli}, which are randomised to avoid order effects. Each screen allows the participant to listen to each audio sample as many times as they need, and select three adjectives from a total of 36 as listed in Table \ref{tab:adjlist}. Adjectives were presented in 4 columns of 9 words, also randomised in each trial to avoid order effects. Each screen also allows for an optional free-text response where extra adjectives or comments can be added.

\begin{table}
\centering
\caption{List of adjectives presented to participants.}
\begin{tabular}{llllllllll} 
\cline{1-2}
\textbf{positive}   & \begin{tabular}[c]{@{}l@{}}charismatic, charming, confident, easygoing, \\ friendly, humorous, intelligent, intimate, \\ kind, playful, pleasant, sincere, warm\end{tabular}  &  &  &  &  &  &  &  &   \\ 
\cline{1-2}
\textbf{negative}   & \begin{tabular}[c]{@{}l@{}}annoying, anxious, awkward, cold, creepy, \\ cringe, fake, forceful, hostile, indifferent, \\ intimidating, unintelligent, unpleasant\end{tabular} &  &  &  &  &  &  &  &   \\ 
\cline{1-2}
\textbf{dominant}   & confident, dominant, forceful, intense,
  serious                                                                                                                             &  &  &  &  &  &  &  &   \\ 
\cline{1-2}
\textbf{submissive} & anxious, awkward, shy, submissive, timid                                                                                                                                      &  &  &  &  &  &  &  &   \\ 
\cline{1-2}
\textbf{sexual}     & exotic, flirty, intimate, seductive,
  sensual                                                                                                                                &  &  &  &  &  &  &  &   \\
\cline{1-2}
\end{tabular}
\label{tab:adjlist}
\vspace{-1em}
\end{table}

We assigned 1-2 labels to each adjective, and ensured that participants saw equal numbers of words in the positive/negative and dominant/submissive sets.
Our classifications were verified by referring to \cite{warriner2013norms}, such that words appearing in our positive/negative lists had scores above/below average for `valence' and dominant/submissive words had above/below average scores for `dominance', respectively. `Sexualised' adjectives had high scores on all three dimensions (valence, arousal and dominance). The word list was refined over several separate pilot studies: we removed particularly ambiguous words and added words that pilot participants had entered in the free text field.
At the end of the study, participants were asked to provide an overall impression of the voices they had listened to, and what they think the study was about. We also asked participant demographic questions such as age, gender (single answer: man, woman, non-binary, agender, prefer not to say, free-text response), attraction (multiple choice, with the same options as in the gender question), ethnicity (single answer), and regular languages used (free-text response).

\subsection{Participants}

Participants are recruited using the online platform Prolific. All participants are current residents of the United States or Canada, with no reported hearing conditions. Our study includes questions about participants' current gender and attraction, and we use this information to sort them into four groups, as listed below. We avoid using descriptions like `straight' or `queer' because we did not ask whether the participants identified with these labels \cite{guyan2022queer, guyan2025rainbow}. We aimed to recruit roughly 30 participants per group; however, Groups 1 and 3 are larger, as they include people who are attracted to multiple genders including their own. Our sample is designed to include representation of non-heterosexual people, who are often excluded in research.
\begin{itemize}
    \item Group 1: women; attracted to women (exclusively or not) (N=40)
    \item Group 2: women; attracted to men only (N=21)
    \item Group 3: men; attracted to men (exclusively or not) (N=34)
    \item Group 4: men; attracted to women only (N=25)
\end{itemize}

\section{Results}

We first address the overall distribution of evaluative judgments for male- and female-coded voices. We fitted generalised linear mixed effects regression models with \textit{nlminbwrap} optimiser, using lme4 in R, to examine the distribution of each group of adjectives with voice gender (female or male), text type (sexualised or Rainbow Passage), participant group (1-4), participant's age (z-scored) as fixed effects and including participant ID as a random effect. The standard alpha level of 0.05 was adopted for all tests. In the entire dataset, we find that male-coded voices are more frequently ascribed \textit{dominant} (\textit{p} = $<$.001) and \textit{positive} (\textit{p} = 0.0027) adjectives.  Female-coded voices are more frequently described using \textit{submissive} (\textit{p} = $<$.001) and \textit{sexualised} (\textit{p} = $<$.001) terms. \textit{Negative} adjectives are distributed roughly equally across male and female voices. 

We also considered content effects of the text spoken in audio samples, comparing ElevenLabs' scripted `flirty' text against excerpts from the Rainbow Passage. Figure \ref{fig:spider_plots_combined} displays the proportion of adjective types for all type of voices. The two sets of sexualised voices with sexualised text receive broadly similar proportions of sexualised adjectives (57\% of all adjectives chosen for female voices, 46\% for male), but when the same voices are paired with Rainbow Passage text, the proportion of sexualised adjectives drops to 36\% for female and 18\% for male voices. This finding suggests that the text content makes a bigger difference to listeners' perceptions of the male voices in the set. One possible explanation is that perceptions of the female voices are also driven by their prosodic and paralinguistic features, like breathiness and sighs, which are absent from the female `presenter' voices.
The same effect is apparent when we examined the most commonly chosen adjectives, across all listeners, for each voice. For 2/3 of the male sexualised voices, listeners' perceptions seem to be content-dependent, as the most used adjectives include \textit{seductive}, \textit{sensual}, and \textit{creepy} when sexualised text is used but change to \textit{confident}, \textit{serious} and \textit{intense} with Rainbow Passage text. For 2/3 of the female equivalents, the most frequently used adjectives are \textit{seductive}, \textit{sensual}, and \textit{intimate} regardless of content.

\begin{figure*}[t]
    \centering
    \begin{subfigure}[b]{0.49\textwidth}
        \centering
        \includegraphics[width=\textwidth, trim = 0 0 0 25]{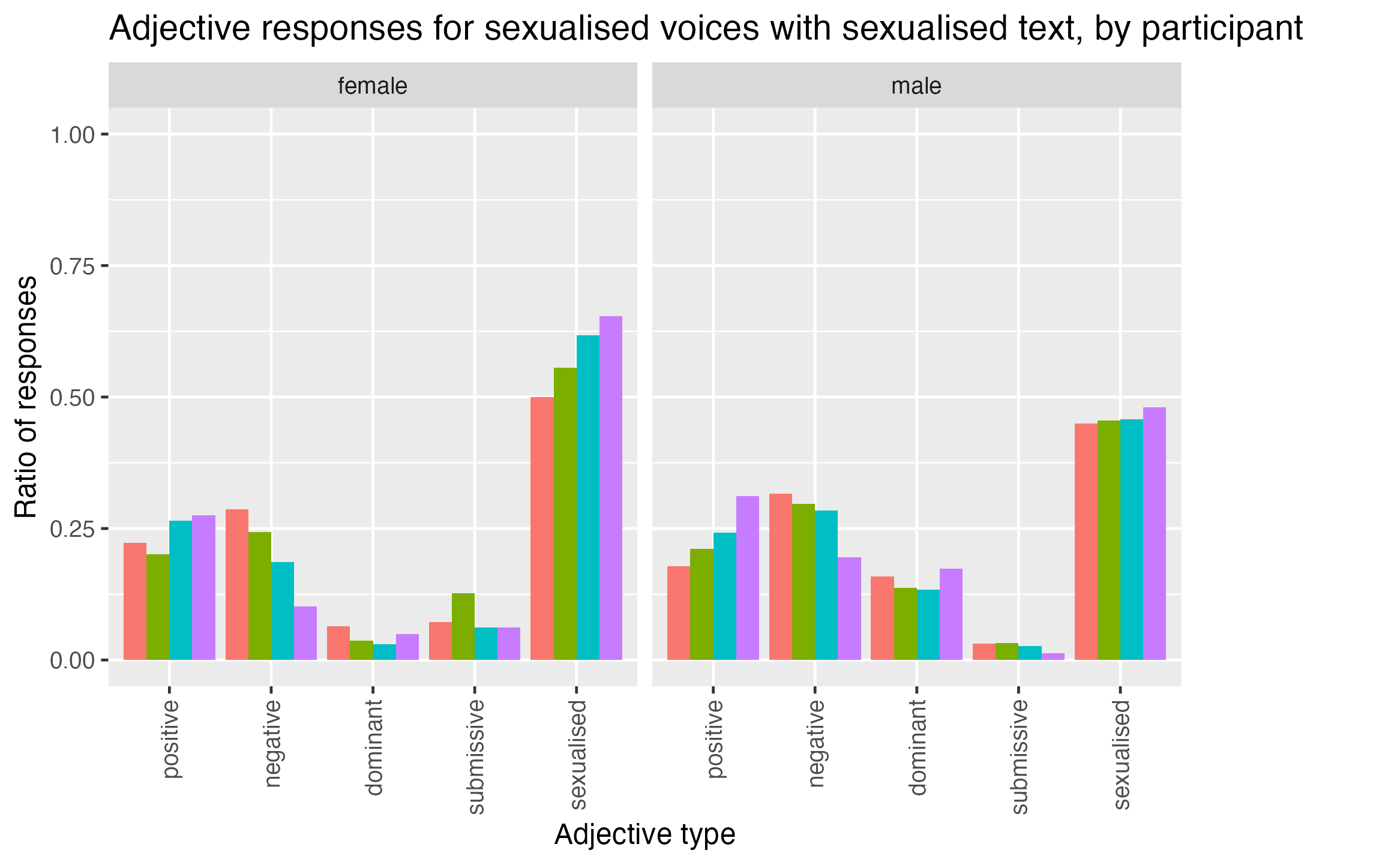}
        \caption{Sexualised voices with sexualised text.}
    \end{subfigure}
    \hfill
    \begin{subfigure}[b]{0.49\textwidth}
        \centering
        \includegraphics[width=\textwidth, trim = 0 0 0 25]{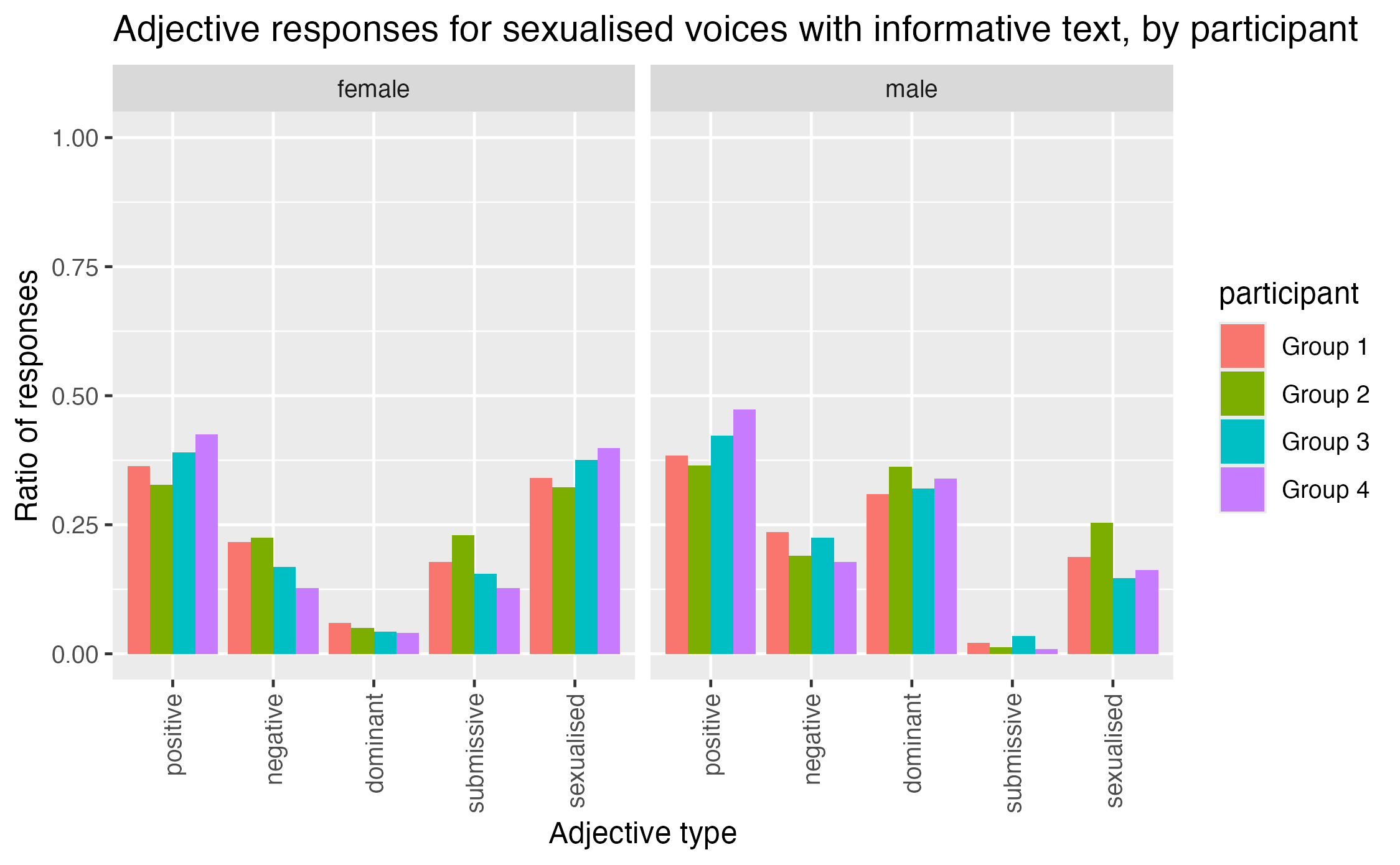}
        \caption{Sexualised voices with informative text.}
        \label{fig:d}
    \end{subfigure}
    \vspace{-1em}
    \caption{Breakdown of adjective types for sexualised voices with different types of text (sexualised/informative) per participant group.}
    \label{fig:results_per_participant}
    \vspace{-1em}
\end{figure*}

Regarding the effects of listeners' own identity, we fitted new linear regression models, using the same package and optimizer, to examine the types of adjectives applied to subsets of the recordings (female and male sexualised voices with sexualised text). These models take participant group (1-4) and age (z-scored) as fixed effects and participant ID as a random effect, and the outcome variable is the adjective type used in each observation (0 or 1).
We observe some statistical differences between Group 4 (men who are attracted to women only) and other groups. Group 4 are more likely than any other to apply sexualised adjectives to sexualised female voices (\textit{p} = 0.0088); less likely to apply negative adjectives to sexualised female voices (\textit{p} = 0.0009); and more likely to choose positive words for sexualised male voices (\textit{p} = 0.0017). Figure \ref{fig:results_per_participant} shows the proportion of adjective categories for each participant type.

Finally, we also calculated mean F0 and speaking rate for each voice in the study to understand general acoustic and prosodic differences between the voices. Both show some differences between the sexualised and the informative voices. Mean F0 is lower for sexualised male voices (70.08 Hz) than informative male voices (110.88 Hz), while female voices show less of a difference (204.73 Hz for sexualised, 211.43 Hz for informative). Speaking rate, defined as the total amount of syllable nuclei divided by duration (in seconds) \cite{de2021praat} shows that sexualised voices are slower on average, regardless of gendering (2.4 for sexualised; 3.8 for informative).

\subsection{Qualitative analysis of participants' comments}
For each audio sample, participants were asked ‘Are there any other adjectives or any comments that you would like to make about the voice?’. 52 participants made one or more optional comments. We analysed 280 comments (151 on female and 129 on male voices), after  excluding comments unrelated to the voices. 53 of these (24 F, 29 M) were positive (e.g., `pleasant', `soothing'), and 143 (81 F, 62 M) negative (e.g., `dull', `uncomfortable to listen to').
Responses like `robotic', `mechanical', `unnatural', `artificial', `AI' were coded as negative, and these account for 35 (24\%) of the negative comments. We coded 84 
comments as neither positive or negative, including a) those describing a voice’s suitability or use case (`like an audiobook'; `like the voice on TikToks'), b) words like `standard', `generic', `masculine' and `feminine', c) comments on speech rate without value judgement (`slow'), or d) descriptions like `sexy', `flirty', `erotic', `porn voice' without expressing a value judgement (24 total; 17 F, 7 M).

Although the number of critical comments was relatively balanced across female (81) and male voices (62), we note some qualitative differences in their content. Multiple negative comments on male voices describe them as `bland', `flat', `dull', `monotonous' or `boring' (n = 16), while others mention `threatening', `predatory' and `rapey' (n = 7). Recurring themes in comments on female voices include `annoying' (3), `jarring' (2), `gross' (2), `harsh' (2), `uncomfortable' (3), and sounding `forced', `trying' or `attempting' (unsuccessfully) to convey positive qualities (6). `Monotonous' and `dull' appear in a smaller proportion for this set (18.6\%) compared with the male voices (35.6\%). A participant in Group 1 wrote: `I hate that I've only marked women's voices as annoying!'. Another, in Group 3, `This sounds more like a teenage boy's idea of what sounds sexy than an actually sensual voice'. Such comments suggest that some listeners perceived the vocal performances as caricatured, male-authored fantasies of femininity. 

\section{Discussion}
We found qualitative and quantitative differences in a diverse group of listeners' reactions to male and female-coded sexualised voices showcased in ElevenLabs' Voice Library. Namely, male voices were more likely to elicit positive adjectives and those relating to high agency or dominance compared to female voices, which more often elicit submissive and sexualised descriptions. 
Comments on the male voices often pointed to either exaggerated dominance or to a lack of affect, while several comments on female voices can reasonably be read as indexing amplified or exaggerated performances of femininity: over-performed, excessively sexualised, and infantilised.
As one participant explains in a comment about their overall impression of the voices, `The more sensual-sounding ones could be uncomfortable to listen to... especially the female ones often felt very stereotyped and made me feel a bit gross'.

The ready-made prompts for these voice personas (see supplementary materials) also reveal how stereotypically masculine and feminine ways of speaking are constructed and linked to value judgments. All three female sexualised voices' prompts mention `breathy'; two of them include the descriptor `honeyed', while male voices are `smooth' and `like aged whiskey'. We also note some unexpected uses of language in prompts, such as `baritone' producing voices that are bass in musical terms. Exoticisation of European accents/identity is apparent: three of the six sexualised voices are described as either French or `Mediterranean', while the informative ones have `neutral American' or `professional American' accents. Several listeners commented that sexualised voices' accents sounded `fake' or inconsistent; one was identified as Irish-accented despite the prompt specifying `a slight French accent' (see also \cite{michel2025not}).

These voices are marketed for use in the creative industries such as video game and film development, presumably supplanting the work of human voice actors like those interviewed in \cite{almeda2025labor} who explained
how workers' rights, safety, and the intimate, personal expression of voice acting were threatened by the framing of endlessly customisable TTS. 
In the same way that video games have historically displayed a very narrow, stereotyped ideal of visual attractiveness for female characters (young, white, thin, with disproportionately large breasts) \cite{lynch2024evidence}, these voices represent a clichéd, heteronormative vocal attractiveness schema.
Given the increasing prevalence of conversational AI, such representations may also influence users’ own speaking styles through processes such as linguistic accommodation and acoustic–prosodic entrainment \cite{szekely2025will}.

\section{Conclusion and future directions}

Personas like `The Parisian temptress', which listeners described as `soooo cringe', `trying to sound sexy but it's just annoying' and `on drugs', are auditory caricatures of a specific performance of femininity. Unlike actual online or phone sex work, though, it  is unclear that any human woman retains agency, control, or profits from this performance.

The 21st century has seen an increase in representation of queer, trans and gender diverse identities in mainstream media. If the voices in this study are representative of ElevenLabs' construction of speakers, they appear to ignore this cultural turn in favour of dated stereotypes. Natural language prompted AI models, on the surface, offer creativity and customisability, but `straight values explicitly and implicitly embedded' in model design can marginalise queer users, as in the case of image generation models \cite{ungless2023stereotypes, taylor2025straightening}. Similarly, the prescribed use of sexualised voices in ElevenLabs' `creative' platform appears to reinforce cisnormative and heteronormative social biases.

The voices examined here represent only one segment of a rapidly expanding market for synthetic speech and AI voice agents. Other high-profile systems, like xAI’s Grok, have been publicly associated with the harassment and exploitation of women \cite{tenbarge2026grok}, while OpenAI's ChatGPT launched its voice chat functionality by apparently deepfaking a famous actress who had refused to license her voice for the project \cite{milmo2024scarlett}. Meanwhile, AI companion platforms increasingly market intimacy, romance, and sexual access as on-demand services \cite{namvarpour2025ai}. As Voice AI continues to proliferate and become normalised, we consider it crucial to examine and understand how artificial voices are adopted and perceived in society.

\section{Acknowledgments}
This study has been approved by the School of Informatics Ethics' Committee at the University of Edinburgh, with reference number 296750. The first authors are supported by the UKRI Centre for Doctoral Training in Natural Language Processing, funded by UKRI (grant EP/S022481/1). Kanhov is supported by the Swedish Research Council (grant number 2024-01012).

\section{Generative AI Use Disclosure}
Generative AI was not used to produce any part of this manuscript. The audio samples used in our listening test were produced using ElevenLabs TTSv3 model, a natural language promptable commercial TTS platform which may be considered a form of Generative AI.

\bibliographystyle{IEEEtran}
\bibliography{mybib}

\end{document}